\documentclass[article]{IEEEtran}
\usepackage[latin1]{inputenc}
\usepackage[T1]{fontenc}
\usepackage{times}
\usepackage{cite}
\usepackage[normalem]{ulem}
\usepackage{pifont}
\usepackage{tikz}
\usepackage{circuitikz}
\usepackage{amsmath}
\usepackage{listings}
 \usetikzlibrary{angles}
 \usetikzlibrary{positioning}
 \usetikzlibrary{decorations}
\usepackage{algorithm}
\usepackage{algpseudocode}
\usepackage{setspace}
 \usetikzlibrary{decorations.pathmorphing}
\usetikzlibrary{shapes.symbols}
\usepackage{tikz-3dplot}
\usepackage{amssymb}
\newcommand{\M}[1]{\mathbf{#1}}
\newcommand{\T}[1]{\mathrm{#1}}
\newcommand{\V}[1]{\boldsymbol{#1}}
\renewcommand{\u}[1]{\boldsymbol{\hat{#1}}}

\usetikzlibrary{shapes.geometric}
\usetikzlibrary{decorations.pathreplacing}

\newcommand{\ie}{\textit{i}.\textit{e}.}
\newcommand{\eg}{\textit{e}.\textit{g}.}
\newcommand{\cf}{\textit{cf}.\,\,}

\title{Characteristic Modes of Frequency-Selective Surfaces and Metasurfaces from S-parameter Data}

\author{Kurt Schab, \IEEEmembership{Member, IEEE}, Frederick W. Chen \IEEEmembership{Student Member, IEEE}, Lukas Jelinek,\\ Miloslav Capek, \IEEEmembership{Senior Member, IEEE}, Johan Lundgren, \IEEEmembership{Member, IEEE}, Mats Gustafsson, \IEEEmembership{Senior Member, IEEE}
\thanks{Manuscript received \today; revised \today. This work was supported by the Czech Science Foundation under project~\mbox{No.~21-19025M}.}
\thanks{K. Schab and F. Chen are with Santa Clara University, Santa Clara, USA (e-mail: \{kschab, zchen7\}@scu.edu).}
\thanks{M. Capek and L. Jelinek are with Czech Technical University in Prague, Czech Republic (e-mails: \{miloslav.capek,lukas.jelinek\}@fel.cvut.cz).}
\thanks{J. Lundgren and M. Gustafsson are with Lund University, Lund, Sweden, (e-mails: \{johan.lundgren, mats.gustafsson\}@eit.lth.se).}
}

\begin{document}

\pagestyle{empty}
\onecolumn

\newpage
\pagestyle{headings}
\twocolumn

\maketitle

\begin{abstract}
  Characteristic modes of arbitrary two-dimensional periodic systems are analyzed using scattering parameter data. This approach bypasses the need for periodic integral equations and allows for characteristic modes to be computed from generic simulation or measurement data. Example calculations demonstrate the efficacy of the method through comparison against a periodic method of moments formulation for a simple, single-layer conducting unit cell. The effect of vertical structure and electrical size on the number of modes is studied and its discrete nature is verified with example calculations. 
  A multiband polarization-selective surface and a beamsteering metasurface are presented as additional examples.
\end{abstract}

\begin{IEEEkeywords}
Antenna theory, eigenvalues and eigenfunctions, frequency-selective surfaces, metasurface, scattering.
\end{IEEEkeywords}

\section{Introduction}

\IEEEPARstart{C}{haracteristic} mode analysis is a widely studied method of decomposing electromagnetic scattering problems into a basis with convenient properties~\cite{apm-1-lau2022characteristic,apm-2-capek2021computational,apm-3-adams2022antenna,apm-4-li2022synthesis,apm-5-manteuffel2022characteristic}. Though an impedance-based formulation of characteristic modes~\cite{Harrington_1971a,HarringtonMautz_ComputationOfCharacteristicModesForConductingBodies,harrington1972characteristic} is most widely used in the literature, scattering-based definitions predate that method~\cite{garbacz1965modal,Garbacz_TCMdissertation} and afford the ability to compute the characteristic modes of arbitrary structures without the requirement of numerical methods based on integral equations~\cite{gustafsson2021unified_part1,gustafsson2021unified_part2,capek2022characteristic}. 
This scattering-based approach to characteristic mode analysis has previously been demonstrated and validated using method of moments (MoM), finite element method (FEM), finite-difference time-domain (FDTD), and hybrid methods~\cite{capek2022characteristic}.

Unlike the vast majority of scatterers studied using characteristic mode analysis, frequency-selective surfaces (FSS) and metasurfaces are typically modeled in a periodic setting~\cite{munk2005frequency} using numerical methods incorporating periodic boundary conditions, \eg, periodic integral equations~\cite{cwik1987scattering}. Previous work on characteristic modes in periodic systems utilizes the impedance-based approach in conjunction with periodic integral equation methods~\cite{Angiulli_2000a,ethier2012antenna, haykir2018characteristic,Haykir_2019a,schab2021sparsity, guo2022calculation, hoffman2023comparison}. Within such a periodic formulation, it becomes apparent that radiating characteristic modes are directly tied to propagating Floquet modes~\cite{schab2021sparsity}. Because existing methods for computing the characteristic modes of periodic systems require surface or volume integral formulations over the unit cell, features such as layered substrates can be included but the implementation and computational complexity can grow rapidly with increasing unit cell inhomogeneity. 
Another, more common strategy is to analyze the characteristic modes of a unit cell in free space before studying its scattering behavior within a periodic lattice~\cite{dicandia2022design,li2020characteristic}. While this approach has been demonstrated as effective in particular applications, it relies on approximate relationships between the behavior of a unit cell in isolation and its behavior within a periodic setting where inter-element coupling may be significant.

In this work, we adopt a scattering formulation of characteristic modes to the study of periodic systems. The added flexibility of this approach enables the study of arbitrary unit cells using a variety of numerical methods, including FEM. The scattering formulation requires only the calculation of S-parameter data, making the technique straightforward to implement using measured data or any solver capable of calculating the S-parameters of a single unit cell within a periodic setting.
  
\section{Scattering-based Periodic Characteristic Modes}
\label{sec:scattering-modes}

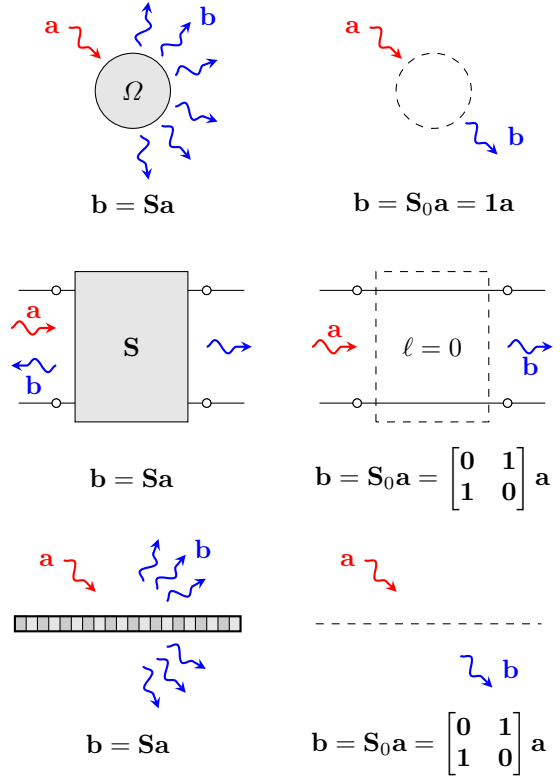
\begin{figure}
  \centering
  \begin{circuitikz}[photon/.style={decorate,decoration={snake,post length=2mm}}]
    
    \begin{scope}[shift={(0,0)}]
        \draw[fill=black!10] (0,0) circle (0.5) node[pos=0.5] {$\varOmega$};
        \node at (0,-1.5) {$\M{b} = \M{S}\M{a}$};

\def\theta{45}
\draw[-stealth,photon,segment length=10pt,thick,red] ({-1.2* cos(\theta)},{1.2*sin(\theta)}) node[left] {$\M{a}$} -- ({-0.6* cos(\theta)},{0.6*sin(\theta)});

\foreach \theta in {20,50,80}{
\draw[-stealth,photon,segment length=10pt,thick,blue] ({0.6* cos(\theta)},{0.6*sin(\theta)}) -- ({1.2* cos(\theta)},{1.2*sin(\theta)}) ;
\draw[-stealth,photon,segment length=10pt,thick,blue] ({0.6* cos(\theta)},{-0.6*sin(\theta)}) -- ({1.2* cos(\theta)},{-1.2*sin(\theta)}) ;}

\node[blue] at (1,1) {$\M{b}$};

    \end{scope}

    \begin{scope}[shift={(4,0)}]
        \draw[dashed] (0,0) circle (0.5);
        \node at (0,-1.5) {$\M{b} = \M{S}_0\M{a} = \M{1}\M{a}$};

        \def\theta{45}
\draw[-stealth,photon,segment length=10pt,thick,red] ({-1.2* cos(\theta)},{1.2*sin(\theta)}) node[left] {$\M{a}$} -- ({-0.6* cos(\theta)},{0.6*sin(\theta)});

\foreach \theta in {45}{
\draw[-stealth,photon,segment length=10pt,thick,blue] ({0.6* cos(\theta)},{-0.6*sin(\theta)}) -- ({1.2* cos(\theta)},{-1.2*sin(\theta)}) node[above right] {$\M{b}$};}
    \end{scope}
    \end{circuitikz}
    ~\\
    \vspace{4ex}
        \begin{circuitikz}[photon/.style={decorate,decoration={snake,post length=2mm}}]
           \begin{scope}[shift={(0,0)}]
        \draw (-0.5,0) to[short,-o] (0,0) to[short,o-o] (2,0) to[short,o-] (2.5,0);
        \draw (-0.5,1.5) to[short,-o] (0,1.5) to[short,o-o] (2,1.5) to[short,o-] (2.5,1.5);
        \draw[fill=black!10] (0.25,-0.25) rectangle (1.75,1.75) node[pos=0.5] {$\M{S}$};
        \node at (1,-1) {$\M{b} = \M{S}\M{a}$};

        \draw[-stealth,photon,segment length=10pt,thick,red] (-0.6,1)  -- node[above] {$\M{a}$}(0,1);
         \draw[-stealth,photon,segment length=10pt,thick,blue] (0,0.5)  -- node[below] {$\M{b}$} (-0.6,0.5) ;
          \draw[-stealth,photon,segment length=10pt,thick,blue] (2,0.75)  -- (2.6,0.75);
    \end{scope}
    
    \begin{scope}[shift={(4,0)}]
        \draw (-0.5,0) to[short,-o] (0,0) to[short,o-o] (2,0) to[short,o-] (2.5,0);
        \draw (-0.5,1.5) to[short,-o] (0,1.5) to[short,o-o] (2,1.5) to[short,o-] (2.5,1.5);
        \draw[dashed] (0.25,-0.25) rectangle (1.75,1.75) node[pos=0.5] {$\ell = 0$};
        \node at (1,-1) {$\M{b} = \M{S}_0\M{a} = \begin{bmatrix}
            \M{0} & \M{1} \\ 
            \M{1} & \M{0}
        \end{bmatrix}\M{a}$};

        \draw[-stealth,photon,segment length=10pt,thick,red] (-0.6,0.75)  --node[above] {$\M{a}$} (0,0.75);

          \draw[-stealth,photon,segment length=10pt,thick,blue] (2,0.75)  --  node[below] {$\M{b}$}(2.6,0.75) ;
    \end{scope}
    \end{circuitikz}
    ~\\
    \vspace{1ex}
\begin{tikzpicture}[photon/.style={decorate,decoration={snake,post length=2mm}}]

\begin{scope}[shift={(0,0)}]
\draw[dashed] (-1.5,0) -- (1.5,0);

\draw[fill=black!10,draw=none] (-1.5,-0.1) rectangle (1.5,0.1);

\foreach \x in {-1.5,-1.2,...,1.35}{
\draw[fill=black!20] ({\x},-0.1) rectangle ({\x+0.15},0.1);}
\draw[fill=none,thick] (-1.5,-0.1) rectangle (1.5,0.1);

\def\theta{45}
\draw[-stealth,photon,segment length=10pt,thick,red] ({-1.2* cos(\theta)},{1.2*sin(\theta)}) node[left] {$\M{a}$} -- ({-0.6* cos(\theta)},{0.6*sin(\theta)}); 

\foreach \theta in {30,50,70}{
\draw[-stealth,photon,segment length=10pt,thick,blue] ({0.6* cos(\theta)},{0.6*sin(\theta)}) -- ({1.2* cos(\theta)},{1.2*sin(\theta)}) ;
\draw[-stealth,photon,segment length=10pt,thick,blue] ({0.6* cos(\theta)},{-0.6*sin(\theta)}) -- ({1.2* cos(\theta)},{-1.2*sin(\theta)}) ;}

\node[blue] at (1,1) {$\M{b}$};

\node at (0,-1.6) {$\M{b} = \M{S}\M{a}$};

\end{scope}

\begin{scope}[shift={(4,0)}]
\draw[dashed] (-1.5,0) -- (1.5,0);

\def\theta{45}
\draw[-stealth,photon,segment length=10pt,thick,red] ({-1.2* cos(\theta)},{1.2*sin(\theta)}) node[left] {$\M{a}$} -- ({-0.6* cos(\theta)},{0.6*sin(\theta)}); 

\foreach \theta in {45}{
\draw[-stealth,photon,segment length=10pt,thick,blue] ({0.6* cos(\theta)},{-0.6*sin(\theta)}) -- ({1.2* cos(\theta)},{-1.2*sin(\theta)}) node[above right] {$\M{b}$};}

\node at (0,-1.6) {$\M{b} = \M{S}_0\M{a} = \begin{bmatrix}
            \M{0} & \M{1} \\ 
            \M{1} & \M{0}
        \end{bmatrix}\M{a}$};

\end{scope}

    \end{tikzpicture}
  \caption{(top left) Scattering by an object $\varOmega$ in free space and (top right) the corresponding background case. (middle left) Scattering by a two-port network and (middle right) the choice of a zero-length through connection as the background case. (bottom left) Scattering between Floquet harmonics by a periodic surface and (bottom right) the corresponding background case of uninterrupted plane wave propagation.}
  \label{fig:background}
\end{figure}

The scattering-based eigenvalue problem used to compute characteristic modes for an object in free space reads~\cite{gustafsson2021unified_part1}
\begin{equation}
  \M{S}\M{a}_n = (2t_n + 1)\M{a}_n,
  \label{eq:sep}
\end{equation}
where $\M{S}$ is the scattering matrix mapping incoming to outgoing waves~\cite{kristensson2016scattering}, shown schematically in the top left panel of Fig.~\ref{fig:background}. In this eigenvalue problem, the vectors $\M{a}_n$ have the dual interpretation of incoming wave coefficients (\ie, characteristic excitations) and scattered wave coefficients (\ie, characteristic far fields), both in an appropriate basis~\cite{garbacz1965modal,gustafsson2021unified_part1,capek2022characteristic}. Following conventions from microwave circuit theory, the wave coefficients are normalized to have units of $\T{W}^{1/2}$, see Appendix~\ref{app:equiv} for details.

The values~ $t_n$ are the characteristic mode eigenvalues associated with the transition matrix, which maps regular waves to outgoing waves~\cite{kristensson2016scattering}. The absolute value of the eigenvalue $t_n$ is equal to modal significance~\cite{gustafsson2021unified_part1}, and that quantity is used throughout the remainder of this paper, as opposed to the eigenvalues $s_n = 2 t_n + 1$ in \eqref{eq:sep}, which exhibit unit modulus due to the unitarity of the matrix $\M{S}$.

For lossless scatterers, equivalence between the scattering formulation in \eqref{eq:sep} and formulations based on several integral equation methods is demonstrated in \cite{gustafsson2021unified_part1}. Interpretation of characteristic modes for lossy structures is nuanced, and the equivalence of scattering and impedance formulations depends on selected orthogonality properties~\cite{harrington1972characteristic,gustafsson2021unified_part2}. For this reason, we consider problems involving only lossless scatterers and lossless background media.

By virtue of the assumption that the obstacle being studied in \eqref{eq:sep} exists ``in free space'', the scattering matrix with no object present is an identity matrix~\cite{Garbacz_TCMdissertation}. Let this ``background'' scattering matrix, sketched in the top right panel of Fig.~\ref{fig:background}, be denoted as $\M{S}_0$. For free-space scattering problems, the eigenvalue problem in \eqref{eq:sep} can therefore be written as ($\M{S}_0$ is an identity matrix)
\begin{equation}
  \M{S}\M{a}_n = (2t_n + 1)\M{S}_0\M{a}_n,
  \label{eq:gep-ss0}
\end{equation}
gaining the slightly different interpretation of finding incident field configurations $\M{a}_n$ which produce the same scattered fields via the total and background scattering matrices, to within a complex scaling factor $2t_n + 1$. Equivalence of \eqref{eq:sep} and \eqref{eq:gep-ss0} for free space problems suggests that the form of \eqref{eq:gep-ss0} may be the appropriate choice when studying problems with non-trivial background scattering matrices, such as the situation encountered in the analysis of periodic systems. This generalization is supported by further analysis appearing later in this section and Appendix~\ref{app:equiv} showing the equivalence of \eqref{eq:gep-ss0} with standard impedance-based methods of computing characteristic modes.

Unlike the free-space scattering scenario, the scattering by a periodic system is closer to that of the multi-port network sketched in the middle row of Fig.~\ref{fig:background}. Consider the special case of a two-port system consisting of two multi-mode measurement ports which, in the absence of any other scattering network, are connected by a lossless, zero-length, perfectly matched transmission line. In this case, the background scattering matrix reads
\begin{equation}
  \M{S}_0 = \begin{bmatrix}
    \M{0} & \M{1}\\
    \M{1} & \M{0}
  \end{bmatrix},
  \label{eq:s0}
\end{equation}
where $\M{1}$ denotes an identity matrix. This choice of background problem in two-port networks is particularly well-suited to the analysis of planar\footnote{Throughout this work, the term \textit{planar periodic system} refers to an arbitrary three-dimensional unit cell infinitely repeated in two dimensions.} periodic systems (\eg{}, metasurfaces or FSS) with S-parameters de-embedded to a common reference plane, as sketched in the bottom row of Fig.~\ref{fig:background}. There the S-parameters of the system represent scattering into propagating Floquet modes above and below the periodic surface.

By the unitary nature of this background matrix, we can rewrite the eigenvalue problem in \eqref{eq:gep-ss0} as
\begin{equation}
  \tilde{\M{S}}\M{a}_n = 2t_n \M{a}_n.
  \label{eq:new-ep-st}
\end{equation}
where
\begin{equation}
  \tilde{\M{S}} = \M{S}_0^\T{H}\M{S} - \M{1}.
  \label{eq:st-new}
\end{equation}

The matrix $\tilde{\M{S}}$ has the physical interpretation of mapping incident positive- and negative-going waves onto the scattered component of the total field in those directions. 
The physical interpretation and motivation for this background matrix rests on the inherent connection between the upper and lower half-spaces when the periodic structure is not present.

Leveraging this interpretation, it can be shown that the eigenvalues $t_n$ in \eqref{eq:new-ep-st} are explicitly related to those obtained by decomposition of an integral impedance operator in the usual form of characteristic modes, see Appendix~\ref{app:equiv},
\begin{equation}
  \M{X}\M{I}_n = \lambda_n\M{R}\M{I}_n,
  \label{eq:gep-z}
\end{equation}
with the relation
\begin{equation}
  t_n = -\frac{1}{1+\T{j}\lambda_n}.
  \label{eq:lambda-t}
\end{equation}
In showing this equivalence, the periodic impedance operator $\M{Z} = \M{R}+\T{j}\M{X}$ represents free-space scattering by currents induced within the periodic environment for a prescribed angle of incidence (phase shift per unit cell). Because of their connection to radiated and reactive power~\cite{Harrington_1971a}, the matrices $\M{R}$ and $\M{X}$ are referred to as the radiation and reactance matrices, respectively. For objects in free space, the transpose and Hermitian symmetry of these matrices guarantees that characteristic currents~$\M{I}_n$ and excitations $\M{a}_n$ are real-valued. In the periodic case with oblique incidence, these matrices are Hermitian symmetric, and the eigenvectors of either characteristic mode eigenvalue problem are no longer real-valued.

Inclusion of background media (\eg, material present when the designable portion of the structure is removed) requires the adoption of a problem-specific Green's function or, equivalently, selection of an appropriate background matrix $\M{S}_0$ representing scattering by the fixed background media. For example, in problems involving the design of patterned conducting elements on a fixed structural substrate, the matrix $\M{S}_0$ may be defined by the scattering behavior of the substrate alone, while the matrix $\M{S}$ describes the complete system including both the substrate and conducting elements. Definition of the background problem, and thus the matrix $\M{S}_0$ is not unique and can be selected to synthesize modes with different interpretations and properties. This is analogous to the definition of controllable and uncontrollable regions in the formulation of substructure characteristic modes~\cite{ethier2012sub}. For simplicity, throughout this paper, we assume a free-space background scattering matrix defined by~\eqref{eq:s0}.

The above analysis demonstrates that the characteristic modes of an arbitrary periodic system can be computed using measurements of two sets of S-parameters: $\M{S}$ (structure present) and $\M{S}_0$ (background measurement, no structure present). Deembedding both sets of S-parameters to a common reference plane makes for a simpler connection to the zero-length, matched, transmission line background case in the center-right panel of Fig.~\ref{fig:background}~, though it is not necessary. Should the reference planes be shifted~ (embedded or de-embedded) via propagation through lossless transmission lines, the two pairs of S-parameters become~\cite[\S 4.3]{pozar2011microwave}
\begin{equation}
  \M{S}' = \M{\Phi}\M{S}\M{\Phi}\quad\T{and}\quad \M{S}_0' = \M{\Phi}\M{S}_0\M{\Phi},
\end{equation}
where $\M{\Phi}$ is a diagonal matrix of complex exponentials. Substitution of these parameters into \eqref{eq:new-ep-st} and \eqref{eq:st-new} leads to identical eigenvalues $t_n$ and eigenvectors $\M{\Phi}\M{a}_n$ embedded to the new reference planes. Hence the characteristic modes can be computed using two sets of S-parameters with arbitrary reference planes, so long as the reference planes are unchanged between the two measurements. Note that this behavior is predicated on the fact that the scattering matrices $\M{S}$ and $\M{S}_0$ contain only propagating modes, which is consistent with the equivalence with impedance-based characteristic modes demonstrated in Appendix~\ref{app:equiv}.

\subsection{Number of radiating CMs}
The number of propagating Floquet harmonics is determined by the size of the unit cell, the Floquet mode index, and the specular wavenumber (phase shift per unit cell). For rectangular unit cells of dimension $T_x\times T_y$, the $\gamma = \{u,v\}$ Floquet harmonic is propagating if its corresponding longitudinal wavenumber $k_{{\gamma}z}$ is real~\cite{cwik1987scattering}, that is, when
\begin{equation}
  k^2 \geq \left(k_{\T{i}x}+\frac{2\pi u}{T_x}\right)^2 + \left(k_{\T{i}y}+\frac{2\pi v}{T_y}\right)^2.
  \label{eq:propagating}
\end{equation}
For low frequencies and near-normal incidence, only the specular $\gamma = \{0,0\}$ mode propagates, leading to a scattering matrix of dimension four\footnote{The total number arises from the two ports, two polarizations, and one propagating Floquet harmonic.}. Increasing frequency or the incident wavevector components can enable higher-order Floquet harmonics to propagate, associated with the onset of grating lobes. Regardless of the frequency or incident wavevector, the condition in \eqref{eq:propagating} determines \emph{a priori} the number of propagating Floquet modes $K$, before any electromagnetic modeling or measurement takes place.

Devices consisting of a single, patterned, two-dimensional current-supporting sheet cannot produce radiation selectively in the $\pm z$ directions, meaning the matrix $\tilde{\M{S}}$ has a maximum rank of $2K$. In contrast, devices supporting currents which vary in the $z$ direction (\eg, dielectrics of finite thickness, multiple conducting sheets) allow for directional scattering, leading to a scattering matrix with a maximum rank of $4K$. 

The rank-limited nature of the scattering matrix $\M{S}$ and the radiation matrix $\M{R}$ leads to a fixed number of radiating characteristic modes. These characteristic modes have finite eigenvalues $\lambda_n$, corresponding to non-zero eigenvalues $t_n$. Sums over the small set of radiating characteristic modes exactly reconstruct the total scattered fields by the unit cell.

These trends were previously observed in impedance-based approaches to characteristic modes~\cite{schab2021sparsity}, but in the scattering-based approach, the underlying eigenvalue problem itself is reduced to the order of the number of radiating Floquet harmonics. This is markedly different from the impedance-based approach, where the matrices involved in the characteristic mode eigenvalue problem retain a high number of spatial degrees of freedom (corresponding to the number of basis functions) regardless of the sparsity of radiating characteristic modes. Hence, the \textit{a priori} truncation of the scattering matrices $\M{S}$ and $\M{S}_0$ to include only propagating modes is again justified by equivalent sparsity produced by matrix systems of much larger dimensions.

\subsection{Decoupling of both temporal and spatial frequencies}

In the preceding analysis, the scattering operator $\M{S}$ and impedance operator $\M{Z}$ are calculated for specific temporal frequency $\omega$ and incident wave vector (spatial frequency) $\V{k}_\T{i}$. Hence the characteristic modes produced by either formulation depend on both of these parameters. Despite the additional dependence on spatial wavenumber, this does not contradict the oft-cited feature that characteristic modes are independent of any prescribed excitation. Rather, it further decouples characteristic modes and the excitations which may induce them into subsets of finer granularity than exists in the finite scatterer case. 

Consider a scatterer of finite extent characterized by the impedance matrix $\M{Z}(\omega)$. The frequency dependence of the impedance matrix is tied to the assumption that any excitation applied to the system carries a time dependence $\T{e}^{\T{j}\omega t}$. Hence the frequency-dependent characteristic modes can be interpreted as the modes that are excitable only by incident fields sharing the same temporal frequency $\omega$. For excitations comprised of multiple temporal frequencies, the union of characteristic modes over those multiple frequencies must be considered, though only sets of modes and excitations sharing the same temporal frequencies need to be considered in expanding a total solution into characteristic modes.

This decoupling of modes and their corresponding excitations has added dimensionality in the case of periodic scatterers analyzed using Floquet methods. There, operators have an explicit dependence on both temporal and spatial frequencies, \eg, the impedance matrix takes the form $\M{Z}(\omega,\V{k}_\T{i})$. Hence the characteristic modes produced by such an impedance 
 or scattering matrix can only be excited by incident fields sharing the same temporal and spatial frequencies. 
For general incident fields with multiple temporal and spatial frequencies, the union of corresponding sets of characteristic modes must be considered. As in the simpler case of the finite scatterer, however, the expansion of any total solution into characteristic modes is block diagonalized by temporal and spatial frequencies.
 
In summary, the property of ``excitation independence'' of characteristic modes carries over into the periodic analysis presented in this work. However, closer examination shows that, even in the case of finite scatterers, this independence is only superficial since there exists an implicit restriction that characteristic modes share the temporal frequency of a selected excitation. In the case of periodic scatterers, this restriction is extended to spatial frequencies as well. In both scenarios, the union of sets of characteristic modes spanning multiple frequencies (or a continuum of frequencies) can be constructed when considering more complex incident fields. However, it should be stressed that coupling between excitations and induced modes occurs only when both share the same temporal and spatial frequency characteristics, greatly simplifying the analysis of systems under complex excitation.

\subsection{Reconstruction of modal currents and fields}
\label{sec:reconstructing-currents}

In the scattering-based formulation~\eqref{eq:new-ep-st}, the characteristic mode eigenvector $\M{a}_n$ represents a modal excitation consisting of incident plane waves from all directions, Floquet harmonics, and polarizations. Applying this excitation leads to scattered fields of the form $t_n\M{a}_n$, \ie, maintaining the same shape as the excitation. Additionally, fields or currents induced by a modal excitation can be taken as alternative representations of characteristic modes. For example, applying a modal excitation~$\M{a}_n$ to a unit cell containing a PEC scatterer induces a surface current distribution which, when represented in an appropriate basis, corresponds to the characteristic mode current distribution~$\M{I}_n$ from \eqref{eq:gep-z}. This process is demonstrated in an example in Sec.~\ref{sec:ex-equiv}.

\subsection{Expansion of S-parameters into CMs}

Collecting the eigenvectors $\M{a}_n$ as the columns of a matrix~$\M{Q}$ and constructing a diagonal matrix $\M{\Lambda}$ containing the values~$2t_n$ allows for \eqref{eq:gep-ss0} to be rewritten as
\begin{equation}
  \M{S}\M{Q} = \M{S}_0\M{Q}(\M{\Lambda}+\M{1}).
\end{equation}
Because both matrices $\M{S}$ and $\M{S}_0$ are unitary under the assumption of losslessness used throughout this paper, the matrix $\M{Q}$ is also unitary~\cite[Ch. 10]{roman2005advanced}.
Rearranging the above expression,~ the S-parameters of the system can be written in terms of the characteristic modes as
\begin{equation}
  \M{S} = \M{S}_0\M{Q}\M{\Lambda}\M{Q}^\T{H} + \M{S}_0,
  \label{eq:exp-1}
\end{equation}
For this particular definition of background problem, multiplication with the matrix $\M{S}_0$ on the right-hand side~ serves only to reorder the entries in a vector or matrix. Let the effect of this reordering on the eigenvector matrix be denoted $\M{S}_0\M{Q} = \widehat{\M{Q}}$ with the columns of $\widehat{\M{Q}}$ being the reordered vectors $\hat{\M{a}}_n$. With this notation, the individual entries of the S-parameter matrix can be written as a sum of modal contributions
\begin{equation}
  S_{ij} = \sum_n S_{ij}^n + S^{\phantom{n}}_{0,ij},
  \label{eq:exp-2}
\end{equation}
where
\begin{equation}
  S_{ij}^n = 2\hat{a}^{\phantom{*}}_{n,i} a_{n,j}^{*} t_n^{\phantom{*}}
  \label{eq:modal-sij}
\end{equation}
are the modal scattering parameters.
Similarly to the expansion of fields and currents in traditional characteristic mode analysis, the presence of the eigenvalue $t_n$ represents a dependence on the modal significance, while the terms $a_{n,j}^{*}$ and $\hat{a}_{n,i}$ represent the projection of each characteristic mode on incoming and outgoing characteristic field configurations. Equivalent to \eqref{eq:exp-1}, the above expressions can be written in matrix form as
\begin{equation}
  \M{S} = \sum_n \M{S}_n + \M{S}_0.
  \label{eq:exp-3}
\end{equation}
where 
\begin{equation}
  \M{S}_n = 2t_n\M{S}_0\M{a}_n\M{a}_n^\T{H}.
  \label{eq:sn}
\end{equation}

The separation of \eqref{eq:exp-2} and \eqref{eq:exp-3} into modal and background contributions is not mandatory. By the unitary nature of the matrix $\M{Q}$, the relation \eqref{eq:exp-3} can be rearranged to yield 
\begin{equation}
  \M{S} = \sum_n (2t^{\phantom{\T{H}}}_n + 1) \M{S}^{\phantom{\T{H}}}_0 \M{a}^{\phantom{\T{H}}}_n \M{a}_n^\T{H},
  \label{eq:exp-4}
\end{equation}
which is mathematically identical, but its interpretation varies slightly from that in \eqref{eq:exp-3}. Whereas in \eqref{eq:exp-3} the scattering parameters consist of a modal sum and a background contribution, in \eqref{eq:exp-4}, the background contribution itself is decomposed into characteristic modes and joined with the other modal terms. 

The choice of using \eqref{eq:exp-3} or \eqref{eq:exp-4} to define modal S-parameters is arbitrary, as the two approaches differ only in how the background scattering parameter is treated. Previous single-mode analysis of periodic systems below the onset of grating lobes in \cite[\S 5.3.2]{ethier2012antenna} utilizes a definition equivalent to~\eqref{eq:exp-4}. In all subsequent examples, we adopt the definition in \eqref{eq:exp-3}. 
Neither definition is necessarily more correct than the other, and it may be advantageous in future work to study whether one approach affords more insight in the analysis or design of particular applications.

\section{Modal tracking}
\label{sec:tracking}

When truncated to include only propagating Floquet harmonics, the dimension of the matrices $\M{S}$ and $\M{S}_0$ change abruptly at the onset of additional propagating Floquet harmonics. Because of this matrix discontinuity, there is no guarantee~\cite{kalaba1981variational} that any modal quantities are continuous across these boundaries in frequency and scan-angle. Hence we focus here only on modal tracking within zones defined by these boundaries. Consider \textit{blocks} of frequencies\footnote{Here ``frequency'' is used to refer generally to the three-dimensional parameter of frequency and two-dimensional transverse incident wavenumber.} which share the same number of propagating Floquet harmonics. Within these blocks, characteristic modes can be tracked using any of a number of standard methods (see \cite[\S V]{capek2022characteristic} for a summary of references). Because the eigenvectors $\M{a}_n$ represent scattered fields, far-field orthogonality can be utilized to carry out far-field tracking when the frequency step size is sufficiently small. For linking the $n^\T{th}$ characteristic mode at the $k^\T{th}$ frequency to its corresponding data point at the $(k+1)^\T{th}$ frequency, far-field tracking consists of simply selecting the mode which maximizes the correlation coefficient magnitude
\begin{equation}
  \left|\rho_{mn}\right| = \left|\left( \M{a}_m^{(k+1)} \right)^\T{H}\M{a}^k_n\right|.
  \label{eq:tracking-correlation}
\end{equation}
In the study of finite structures, this form of tracking is highly effective but its implementation can be complicated by the ``appearance'' of modes at certain frequencies due to shifts in the order of modal significances or in modal eigenvalues crossing the threshold of numerical noise. For the periodic systems considered here, the fact that the number of propagating modes is fixed within each block removes these issues entirely.

\section{Example calculations}

In this section, we consider several examples designed to demonstrate critical features of scattering-based characteristic mode analysis for periodic structures.

\subsection{Equivalency with impedance-based methods}
\label{sec:ex-equiv}
To demonstrate the equivalency of the impedance-based and scattering-based formulations, we analyze the scattering properties of a single-layer PEC FSS using both methods. For this example, a square unit cell has dimension $T$ and contains a rectangular PEC patch of size $0.4T\times0.5T$. For the scattering-based method, the matrix $\tilde{\M{S}}$ is generated using S-parameters computed from a finite element solver (Ansys HFSS~\cite{hfss2021}). Characteristic mode eigenvalues $t_{n}$ are then computed by \eqref{eq:new-ep-st}. For the impedance-based method, the EFIE impedance matrix data are generated using the method of moments \cite{cwik1987scattering, schab2021sparsity}. In this case, characteristic mode eigenvalues $\lambda_{n}$ are computed by \eqref{eq:gep-z} and then converted into the eigenvalues $t_{n}$ via \eqref{eq:lambda-t}. 

\begin{figure}
  \centering
  \includegraphics[width=3.25in]{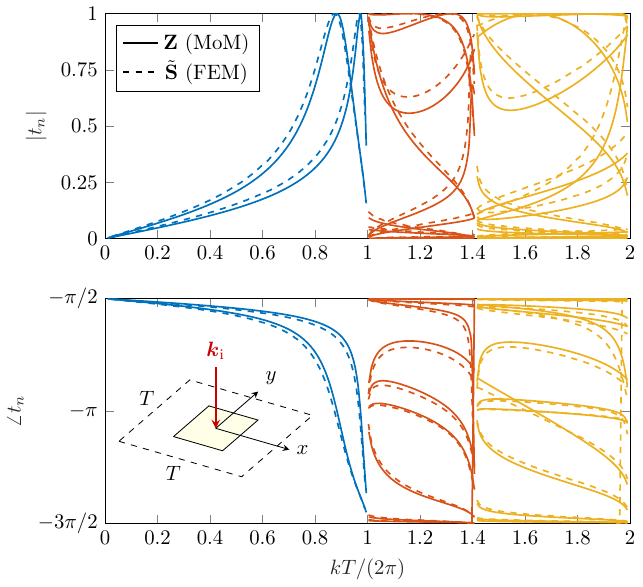}\\
  \caption{Tracked eigenvalue magnitudes $|t_n|$ produced by the scattering-based approach using FEM data (HFSS, dashed lines) and by the MoM impedance formulation (solid lines) for normal incidence on a square unit cell of dimension $T$ containing a $0.4T\times 0.5T$ perfectly conducting patch. 
 Trace colors correspond to blocks of frequencies sharing the same number of propagating Floquet harmonics.}
  \label{fig:equivalency-00}
\end{figure}

\begin{figure}
  \centering
  \includegraphics[width=3.25in]{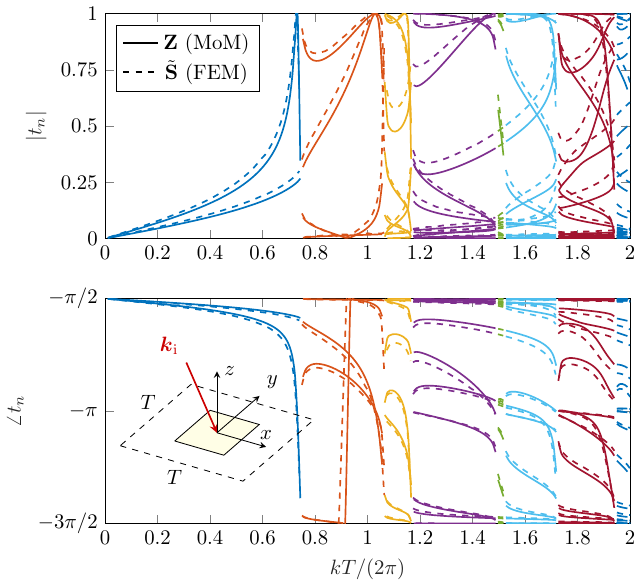}\\
  \caption{Tracked eigenvalue magnitudes $|t_n|$ produced by the scattering-based approach using FEM data (HFSS, dashed lines) and by the MoM impedance formulation (solid lines) for oblique incidence ($\phi = 0^\circ,~\theta=20^\circ$) on a square unit cell of dimension $T$ containing a $0.4T\times 0.5T$ perfectly conducting patch. 
 Trace colors correspond to blocks of frequencies sharing the same number of propagating Floquet harmonics.}
  \label{fig:equivalency-20}
\end{figure}

Figs.~\ref{fig:equivalency-00} and \ref{fig:equivalency-20} show the modal significance $|t_{n}|$ and characteristic angle $\alpha_n = \angle t_n$ over a range of frequencies, normalized as electrical size $T/\lambda = kT/({2\pi})$ for normal ($\theta = 0^\circ,~\phi = 0^\circ$) and oblique ($\theta = 20^\circ,~\phi = 0^\circ$) incidence. The eigenvalues $t_n$ from both methods are tracked using the block-wise correlation method discussed in Sec.~\ref{sec:tracking}. Changes in the trace colors correspond to block boundaries and the onset of additional propagating modes.

As expected, the modal significances generated by both formulations agree across the entire range of frequencies. Small discrepancies in the characteristic angle $\alpha_n$ manifest as larger differences in the modal significance $|t_n|$. The matrix $\tilde{\M{S}}$ was also produced in the method of moments using~\eqref{eq:s-from-z}. Decomposition of that matrix via \eqref{eq:new-ep-st} leads to characteristic mode eigenvalues which are numerically identical to those shown in Figs.~\ref{fig:equivalency-00} and \ref{fig:equivalency-20} produced by the impedance formulation. The observed discrepancies are therefore solely due to different electromagnetic solvers used to compute the MoM and FEM datasets. 

\newcommand{\outarrow}[4]{ \foreach \ang in {0,10,...,360} { \draw[shift={(\ang:1pt)},black,-stealth,ultra thick] (#1,#2) -- (#3,#4);}; \draw[ultra thick,white,-stealth] (#1,#2) -- (#3,#4);}

\begin{figure}
  \centering
  \begin{tikzpicture}
  \node at (0,0) {\includegraphics[width=0.7in]{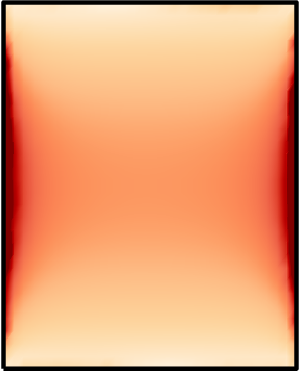}};
  \begin{scope}[shift={(0,0)}]
  \outarrow{0}{0.25}{0}{-0.25}
  \end{scope}
  \node at (2,0) {\includegraphics[width=0.7in]{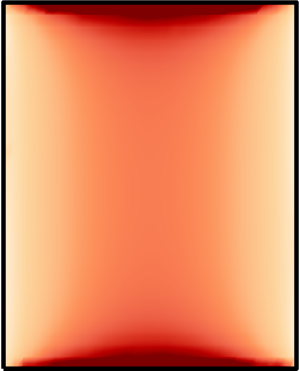}};
  \begin{scope}[shift={(2,0)}]
  \outarrow{0.25}{0}{-0.25}{0}
  \end{scope}
  \node at (4,0) {\includegraphics[width=0.7in]{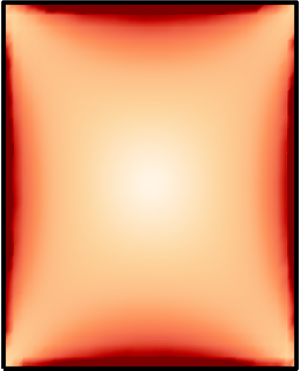}};
  \begin{scope}[shift={(4,0)}]
  \outarrow{-0.5}{-0.25}{-0.5}{0.25}
  \outarrow{0.5}{0.25}{0.5}{-0.25}
  \outarrow{0.25}{0.7}{-0.25}{0.7}
  \outarrow{-0.25}{-0.7}{0.25}{-0.7}
  \end{scope}
  \node at (6,0) {\includegraphics[width=0.7in]{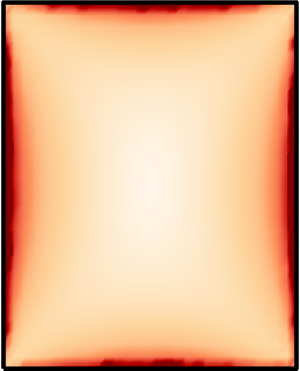}};
  \begin{scope}[shift={(6,0)}]
  \outarrow{-0.5}{-0.25}{-0.5}{0.25}
  \outarrow{0.5}{0.25}{0.5}{-0.25}
  \outarrow{-0.25}{0.7}{0.25}{0.7}
  \outarrow{0.25}{-0.7}{-0.25}{-0.7}
  \end{scope}

  \node[scale=0.8,transform shape,align=center] at (0,-1.55) {$|t_1| = 1.0$\\$\lambda_1 = 0.07$};
  \node[scale=0.8,transform shape,align=center] at (2,-1.55) {$|t_2| = 0.98$\\$\lambda_2 = -0.22$};
  \node[scale=0.8,transform shape,align=center] at (4,-1.55) {$|t_3| = 0.63$\\$\lambda_3 = -1.2$};
  \node[scale=0.8,transform shape,align=center] at (6,-1.55) {$|t_4| = 0.42$\\$\lambda_4 = 2.1$};
  \end{tikzpicture}
  \caption{Characteristic surface current magnitude associated with the four modes with the largest eigenvalue magnitudes $|t_n|$ at the frequency $kT/(2\pi) = 1.2$ calculated using FEM (HFSS) S-parameter simulation of the single-layer unit cell drawn in Fig.~\ref{fig:equivalency-00} under normal incidence illumination. Dark (light) colors denote the maxima (minima) of the normalized surface current density magnitude. Arrows depict the dominant surface current orientations.}
  \label{fig:currents}
\end{figure}
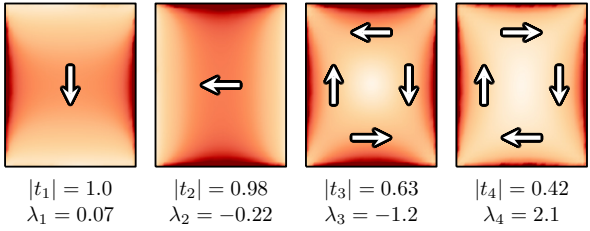

An example of reconstructing characteristic mode current distributions from scattering data is shown in Fig.~\ref{fig:currents}, where we consider the four most significant modes of the single-layer unit cell under normal excitation. The four panels show the modal surface current distribution over the conducting patch computed using FEM (Ansys HFSS~\cite{hfss2021}), where the structure is illuminated with a superposition of plane waves with weights governed by the characteristic mode eigenvector~$\M{a}_n$, see Sec.~\ref{sec:reconstructing-currents}. Bold arrows have been added to facilitate identification of the dominant surface current orientations. Modes 1 and 2 are near resonance and have small positive (inductive) and negative (capacitive) eigenvalues, respectively. Modes 3 and 4 are further away from resonance and exhibit multipole capacitive and loop-like inductive behavior, respectively. It is important to stress that Fig.~\ref{fig:currents} shows only the periodic current distribution on a single unit cell. The fields and interactions between adjacent unit cells also significantly impact the reactive nature of each mode, \ie, their eigenvalues and modal significance. Additionally, the modal currents shown here are the periodic modal currents without the progressive phase shift present in the true current distribution covering the entire periodic surface, see \eqref{eq:jper} in Appendix~\ref{app:equiv} for further details.

\subsection{Vertical structure and the number of propagating modes}

Based on the discussion in Sec.~\ref{sec:scattering-modes}, a structure with only one infinitely thin conducting layer can only scatter in a symmetric fashion above and below the plane of the unit cell. It follows that for such structures the frequency-dependent number of radiating characteristic modes is equal to $2K$, where $K$ is the number of propagating Floquet harmonics at each frequency. In contrast, a two-layer structure is capable of supporting unidirectional scattering, and this additional degree of freedom increases the number of radiating characteristic modes to $4K$. Adding further vertical structure does not afford further scattering diversity, and a structure with three or more layers will also exhibit $4K$ radiating characteristic modes.

\begin{figure}
  \centering
  \includegraphics[width=3in]{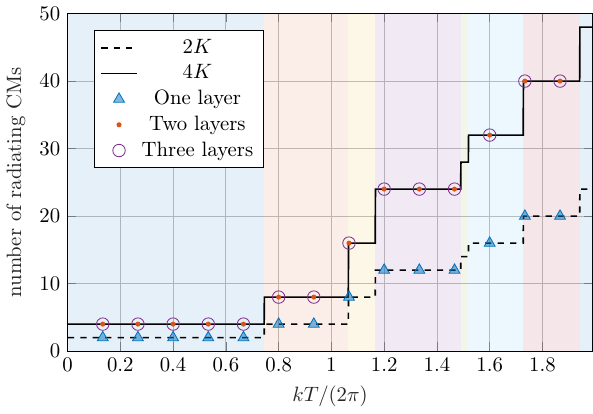}\\
  ~\\
  \tdplotsetmaincoords{60}{30}

\begin{tikzpicture}[tdplot_main_coords,scale=0.9,transform shape,photon/.style={decorate,decoration={snake,post length=2mm}}]

\definecolor{yellow}{rgb}{0.00000,0.44700,0.74100}%
\draw[fill = yellow!10] (-0.4,-0.5,0) -- (0.4,-0.5,0) -- (0.4,0.5,0) -- (-0.4,0.5,0) -- (-0.4,-0.5,0);

\draw[dashed] (-1,-1,0) -- node[below] {$T$} (1,-1,0) -- (1,1,0) -- (-1,1,0) --  node[left,pos=0.3] {$T$~~} (-1,-1,0);

\end{tikzpicture}
~\begin{tikzpicture}[tdplot_main_coords,scale=1,photon/.style={decorate,decoration={snake,post length=2mm}}]
\definecolor{yellow}{rgb}{0.85000,0.32500,0.09800}%

\draw[fill = yellow!10] (-0.4,-0.5,-0.1) -- (0.4,-0.5,-0.1) -- (0.4,0.5,-0.1) -- (-0.4,0.5,-0.1) -- (-0.4,-0.5,-0.1);
\draw[fill = yellow!10] (-0.4,-0.5,0.1) -- (0.4,-0.5,0.1) -- (0.4,0.5,0.1) -- (-0.4,0.5,0.1) -- (-0.4,-0.5,0.1);

\draw[dashed] (-1,-1,0) -- (1,-1,0) -- (1,1,0) -- (-1,1,0) -- (-1,-1,0);
\end{tikzpicture}
~\begin{tikzpicture}[tdplot_main_coords,scale=1,photon/.style={decorate,decoration={snake,post length=2mm}}]

\definecolor{yellow}{rgb}{0.49400,0.18400,0.55600}%

\draw[fill = yellow!10] (-0.4,-0.5,-0.2) -- (0.4,-0.5,-0.2) -- (0.4,0.5,-0.2) -- (-0.4,0.5,-0.2) -- (-0.4,-0.5,-0.2);
\draw[fill = yellow!10] (-0.4,-0.5,0) -- (0.4,-0.5,0) -- (0.4,0.5,0) -- (-0.4,0.5,0) -- (-0.4,-0.5,0);
\draw[fill = yellow!10] (-0.4,-0.5,0.2) -- (0.4,-0.5,0.2) -- (0.4,0.5,0.2) -- (-0.4,0.5,0.2) -- (-0.4,-0.5,0.2);

\draw[dashed] (-1,-1,0) -- (1,-1,0) -- (1,1,0) -- (-1,1,0) -- (-1,-1,0);

\end{tikzpicture}
  \caption{The effect of vertical structure and electrical size on the number of radiating characteristic modes. The frequency-dependent number of radiating characteristic modes for unit cells containing one, two, or three rectangular PEC patches are shown alongside scaled analytic calculations of the number of propagating Floquet harmonics, $K$. Shaded regions correspond to the blocks tracked characteristic modes in Fig.~\ref{fig:equivalency-20}.}
  \label{fig:multilayer}
\end{figure}

To verify the effect of vertical structure on the predictable number of radiating characteristic modes, we examine the set of systems depicted in Fig.~\ref{fig:multilayer} consisting of one, two, or three stacked, rectangular, PEC patches of dimension $0.4T\times 0.5T$ centered within a square unit cell of dimension $T \times T$. In the cases involving two or three patches, adjacent patches are aligned in the $xy$ plane and separated by a vertical distance $\Delta = T/10$ in the $z$ direction. For this example, we consider the incidence angle $(\phi = 0^\circ$, $\theta = 20^\circ)$. The frequency-dependent number of characteristic modes for structures with one, two, and three layers are shown in Fig.~\ref{fig:multilayer}, where perfect alignment with the analytic predictions of either $2K$ or $4K$ is observed.  Results shown in Fig.~\ref{fig:multilayer} were calculated using the impedance formulation~\eqref{eq:gep-z} and data produced by periodic MoM. Identical results were obtained using the scattering formulation~\eqref{eq:new-ep-st} and data from FEM. Note that the single layer markers and $2K$ curve in Fig.~\ref{fig:multilayer} align with the number of traces within each block of Fig.~\ref{fig:equivalency-20}.

\subsection{Analysis of a circular polarization-selective surface (CPSS)}

As a practical example of applying characteristic mode analysis to a more complex design, we consider the CPSS reported in \cite{lundgren2018design}. The CPSS is designed to achieve polarization-selective operation in two bands. In the lower $17.7$--$20.2\,\T{GHz}$ band, the CPSS passes left-hand circular polarization (LHCP) and reflects right-hand circular polarization (RHCP). The opposite behavior occurs in the higher $27.5$--$30.0\,\T{GHz}$ band. The design consists of six metallic meander-like patterns separated by a combination of dielectric substrates, low-permittivity spacers, and bonding layers. For the purposes of characteristic mode analysis, all materials are assumed to be lossless.

\begin{figure}
  \centering
\includegraphics[width=0.45 \textwidth]{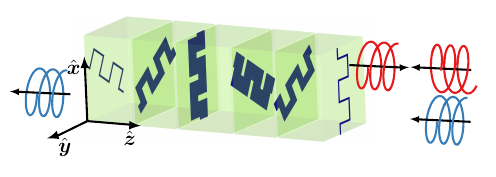}
  \caption{Illustration of the tilted layered metal meander-based dual-band CPSS design~\cite{lundgren2018design}. RHCP (red) and LHCP (blue) waves incident from the negative $z$-direction are reflected and transmitted, respectively, by the periodic structure.}
  \label{fig:CPSSdesign}
\end{figure}

The S-parameters of the CPSS are simulated for normal incidence over a broad bandwidth below the onset of grating lobes ($K = 1$) using CST FEM solver with $60^\circ$ rhombus unit cells~\cite{cst}. The design for one unit cell is illustrated in Fig.~\ref{fig:CPSSdesign}.

\begin{figure}
  \centering
\includegraphics[width=3.25in]{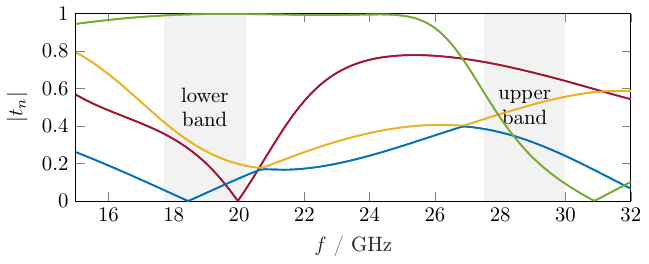}
  \caption{Modal significances $|t_n|$ computed for the CPSS structure shown in Fig.~\ref{fig:CPSSdesign} and described further in \cite{lundgren2018design}. Lower and upper operational bands are highlighted in gray.}
  \label{fig:eigs-cpss}
\end{figure}

Two datasets are collected and stored as $\M{S}$ and $\M{S}_0$, corresponding to the S-parameters with and without the CPSS present in the simulated unit cell, respectively. In defining circular polarizations at each port, a $60^\circ$ offset between Cartesian coordinate systems on either side of the CPSS is used to maintain inverse symmetry of the system. Both sets of S-parameters are de-embedded to a common reference plane. The characteristic modes are then computed via \eqref{eq:new-ep-st} and tracked using the correlation method in \eqref{eq:tracking-correlation}. 
 Modal significances $|t_n|$ for the four radiating characteristic modes on the structure are shown in Fig.~\ref{fig:eigs-cpss}. Modal and background contributions to the S-parameters are then calculated using \eqref{eq:modal-sij}.

\begin{figure}
  \centering  \includegraphics[width=3.25in]{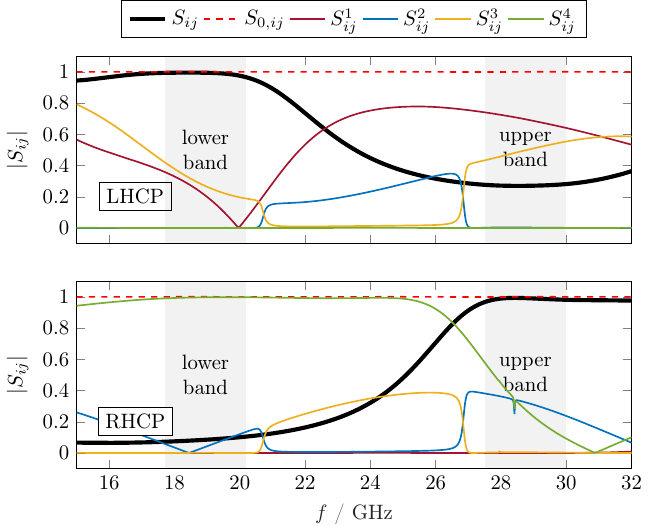}
  \caption{Total (black, thick), background (red, dashed), and modal (thin) transmission scattering parameter magnitude $|S_{ij}|$ for LHCP (top) and RHCP (bottom). Lower (LHCP pass, RHCP stop) and upper (RHCP pass, LHCP stop) operational bands are highlighted in gray.}
  \label{fig:modal-sij}
\end{figure}

\begin{figure}
  \centering  \includegraphics[width=3.15in]{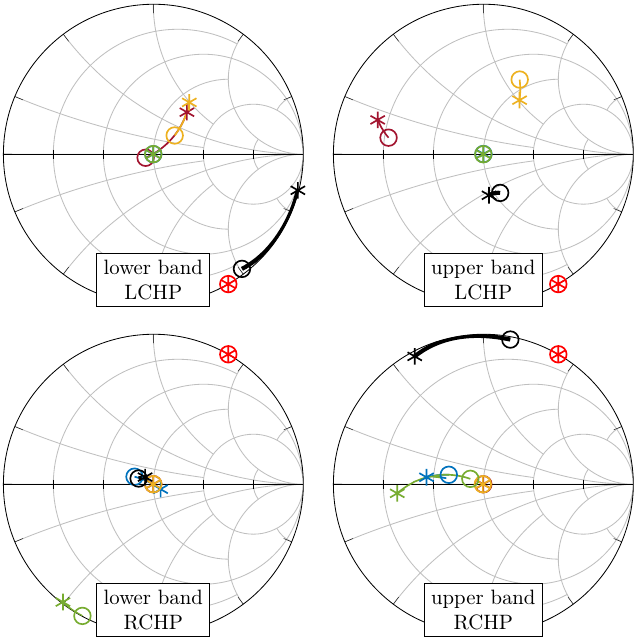}
  \caption{Complex transmission scattering parameter $S_{ij}$ for LHCP (top row) and RHCP (bottom row) in the lower (left column) and upper (right column) operational bands highlighted in Fig.~\ref{fig:modal-sij}. Trace colors and styles follow the legend of Fig.~\ref{fig:modal-sij}. Star and circle markers denote the lowest and highest frequency of each band, respectively.}
  \label{fig:complex-sij}
\end{figure}

LHCP and RHCP transmission parameters (denoted generically as $S_{ij}$) are plotted in Figs.~\ref{fig:modal-sij} and \ref{fig:complex-sij}. In Fig.~\ref{fig:modal-sij}, the S-parameter magnitude (thick, black) and its modal (solid) and background (dashed) components are plotted as functions of frequency. The two operating bands are highlighted with gray shading. Fig.~\ref{fig:complex-sij} shows the complex total, background, and modal S-parameter data over those two bands.

These data offer several points of interpretation regarding the total device behavior in terms of background and modal contributions. In the passband of each polarization (LHCP / lower and RHCP / upper), modal contributions destructively combine with themselves, leaving the transmission dominated by the background transmission parameter -- with a slight modification to the overall transmission phase arising from the weakly excited modes. The contrary occurs in the stopband of each polarization where the modal contributions destructively interact with the background. Low transmission is synthesized by a superposition of one ($S^4$ for RHCP) or two ($S^2$ and $S^3$ for LHCP) modal transmission coefficients which cancel the transmission due to the background. These complex values and combinations can be further seen in Fig.~\ref{fig:complex-sij}.

\subsection{Analysis of a periodic beamsteering metasurface}

The previous CPSS example was designed to operate below the onset of grating lobes with reflection and transmission confined to specular directions. To enhance coupling into non-specular directions above the onset of grating lobes, many designs employ electrically large supercells consisting of multiple dissimilar elements, each tailored for particular scattering phase and polarization characteristics. This periodic metasurface design approach is widespread in the design of microwave and optical devices~\cite{kildishev2013planar}. Large, non-periodic structures can also be designed using this general approach~\cite{minatti2014modulated,faenzi2019metasurface}, but here we focus on the analysis of periodic systems consisting of large, variable element supercells.

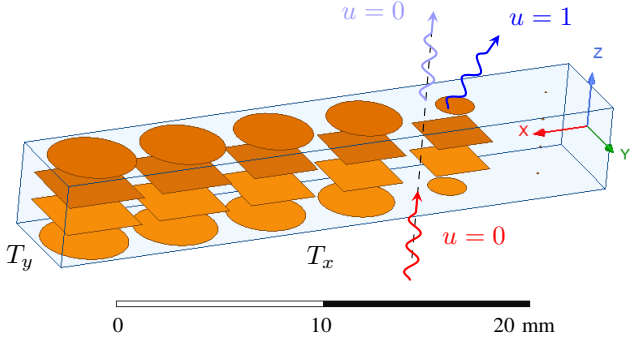
\begin{figure}
  \centering
  \begin{tikzpicture}[photon/.style={decorate,decoration={snake,post length=2mm}}]
    \node at (0,0) {\includegraphics[width=3.5in]{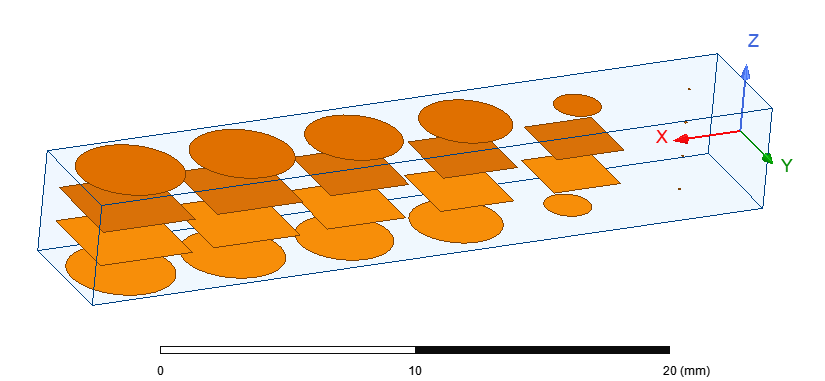}};
    \node at (-4,-1.1) {$T_y$};
    \node at (0,-1.1) {$T_x$};
    \node[fill=white] at (-2.7,-2) {\footnotesize 0};
    \node[fill=white] at (0,-2) {\footnotesize 10};
    \node[fill=white] at (2.7,-2) {\footnotesize 20~mm};
    \def\theta{45};
    \begin{scope}[rotate=-5,shift={(1.3,0.5)}]
    \draw[black,dashed] (0,-1.5) -- (0,1.5);
    \foreach \theta in {60}{
    \draw[-stealth,photon,segment length=10pt,thick,blue] ({0.6* cos(\theta)},{0.6*sin(\theta)}) -- ({1.8* cos(\theta)},{1.8*sin(\theta)}) node[above right] {$u=1$} ;}
    \foreach \theta in {90}{
    \draw[-stealth,photon,segment length=10pt,thick,blue!40] ({0.6* cos(\theta)},{0.6*sin(\theta)}) -- ({1.8* cos(\theta)},{1.8*sin(\theta)}) node[left] {$u=0$~~~};}
    \foreach \theta in {90}{
    \draw[-stealth,photon,segment length=10pt,thick,red] ({1.8* cos(\theta)},-{1.8*sin(\theta)})({1.8* cos(\theta)},-{1.8*sin(\theta)}) -- node[right] {~~$u=0$} ({0.6* cos(\theta)},-{0.6*sin(\theta)});}
    \end{scope}
  \end{tikzpicture}
  \caption{Six-element supercell design from \cite{singh2019controlling} forming a unit cell in the $xy$-plane with dimensions $T_x = $, $T_y = $. The supercell consists of PEC patches (orange) on and within a Taconic TLY-5 substrate (blue, $\epsilon_\T{r} = 2.2$). The size of elements within the supercell increases along the $x$-axis, with the rightmost element being very small with respect the others. The surface converts incoming normally incident plane waves (red, $u = 0$) into a transmitted plane wave propagating in the $\theta = 30^\circ$ direction (blue, $u=1$).}
  \label{fig:beamsteer-geo}
\end{figure}

As an example, we consider the six-element beamsteering supercell reported in \cite{singh2019controlling}. This design, shown in Fig.~\ref{fig:beamsteer-geo}, consists of a supercell containing six unique elements, each constructed from four PEC patches and three dielectric support layers. The design is optimized to convert a normally-incident $20$\,GHz plane wave into a transmitted plane wave exiting the structure at an angle of $\theta = 30^\circ$. For this particular size supercell, the normal and $\theta = 30^\circ$ waves correspond to the $u = 0$ and $u=1$ Floquet harmonics, respectively. 

\begin{figure}
  \centering
  \includegraphics[width=3.25in]{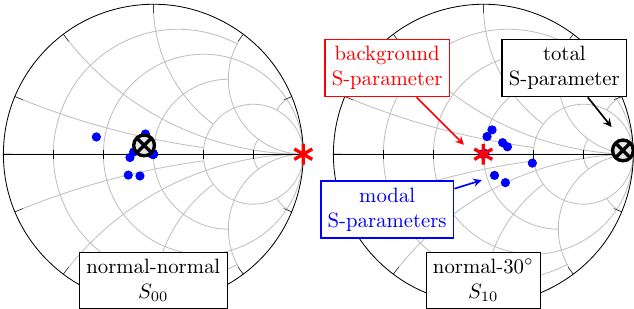}
  \caption{Transmission S-parameters describing normal-normal (left) and normal-$30^\circ$ (right) scattering from the supercell in Fig.~\ref{fig:beamsteer-geo} at 20\,GHz.}
  \label{fig:beamsteer-data}
\end{figure}

Scattering parameters describing normal-normal ($S_{00}$) and normal-$30^\circ$ ($S_{10}$) transmission through the surface are shown in Fig.~\ref{fig:beamsteer-data} at the design frequency of $20$\,GHz. Black markers indicate the total scattering parameters, with near-zero normal-normal transmission and near-unity normal-$30^\circ$ transmission. The background scattering parameters, indicated by red markers, are essentially the opposite, since in the absence of the beamsteering surface the incident normal plane wave propagates directly through the system with no mode conversion or reflection. However, by \eqref{eq:exp-2}, the total scattering parameters are obtained through the sum of the background parameters with the modal scattering parameters, marked in blue. 

In the case of normal-normal transmission, the background S-parameter is cancelled through the sum of many weakly contributing modal S-parameters sitting to the left of the origin. Conversely, the nearly-full normal-$30^\circ$ transmission is produced by the coherent sum of many weakly contributing modal S-parameters sitting to the right of the origin. In this particular example, the total system response contains contributions from many radiating characteristic modes, making the interpretation more challenging than in the sparser CPSS example, \cf Fig.~\ref{fig:complex-sij}. Nevertheless, detailed examination of modal field and current distributions may elucidate how alterations to the structure could be made to alter its performance, \eg, shifting the operating frequency or switching to an alternate transmitted Floquet harmonic, as in the study of finite objects.

\section{Conclusions}

In this work, the characteristic modes of periodic systems, such as frequency- or polarization-selective surfaces, are formulated using a scattering-based approach. The proposed method requires only knowledge of a system's scattering parameters. This lack of dependency on integral equations greatly expands the application of characteristic modes to unit cells consisting of arbitrary materials. The method also allows for the use of measured scattering parameters to compute characteristic mode data, though this is limited to the modal eigenvalues and not field quantities, such as modal currents.

Due to the generally low number of characteristic modes, modal tracking is computationally efficient and straightforward to implement over blocks of frequencies away from the onset of additional grating lobes. Further, the vertical structure of a unit cell plays a crucial role in determining the number of radiating modes. Finally, the expansion of S-parameters into modal contributions again follows the theme of characteristic modes, which is to represent complex system responses as a sum of simpler modal responses. For the selected CPSS example, expansion into characteristic modes illustrates that the transmission in each pass-band is dominated by the background contribution, while each stop-band can be understood as a cancellation of that background contribution by one or two modal terms. Similar features are present in the case of the overmoded beamsteering metasurface, though in that example the modal sums are comprised of many weak contributions from nearly all characteristic modes.

As in the study of finite objects, the impact of lossy materials on the interpretation and utility of characteristic modes must be carefully considered\cite{kuosmanen2022orthogonality,gustafsson2021unified_part2}, especially when the objects under study are specifically designed to dissipate large amounts of incident energy, \eg, absorbing metasurfaces. Adaptation of the proposed method to lossy systems, along with a detailed study of the properties of the resulting characteristic modes, is left as future work. 


\appendices

\section{Equivalence of scattering and impedance-based characteristic modes}
\label{app:equiv}

Equivalence between the scattering- and impedance-based formulations is demonstrated here in four steps. First, the scattered Floquet mode coefficient $b_m$ is written as a product of expansion coefficients $\M{I}$ describing a current density within a unit cell and a matrix $\M{U}_m$ representing projections of Bloch modes onto used basis functions. Second, projection of incident plane wave on the selected basis is shown to be representable in terms of the same operator $\M{U}_m$. The mappings between incident field, induced current density, and scattered fields are combined to relate the matrix $\tilde{\M{S}}$ to an impedance matrix $\M{Z}$. Finally, that relation is used to demonstrate equivalency between the eigenvalue problems in \eqref{eq:st-new} and \eqref{eq:gep-z} along with the eigenvalue conversion \eqref{eq:lambda-t}.

Consider a periodic system supporting an equivalent volumetric current density of the form
\begin{equation}
  \V{J}(x,y,z) = \V{j}(x,y,z)\T{e}^{-\T{j}k_{ix}x}\T{e}^{-\T{j}k_{iy}y}
  \label{eq:jper}
\end{equation}
where $\V{k}_\T{i}$ is an incident wavevector (governing phase shift per unit cell) and $\V{j}(x,y,z)$ is a periodic function that is identical within each unit cell. A rectangular unit cell of dimension $T_x\times T_y$ is considered below without loss of generality. The Fourier components of the periodic current are given by 
\begin{equation}
  \tilde{\V{\jmath}}_\gamma(z) = \frac{1}{T_xT_y}\int \V{j}(x,y,z)\T{e}^{\T{j}\kappa_{\gamma x}x}\T{e}^{\T{j}\kappa_{\gamma y}y}\,\T{d}x\,\T{d}y
  \label{eq:jtilde}
\end{equation}
where
\begin{equation}
  \kappa_{\gamma x} = \frac{2\pi u}{T_x},\quad \kappa_{\gamma y} = \frac{2\pi v}{T_y},
\end{equation}
and $\gamma$ denotes a combination of the Floquet indices $u$ and $v$. The current density~$\V{J}$ is then given by
\begin{equation}
  \V{J} \left( x,y,z \right) = \sum\limits_\gamma \tilde{\V{\jmath}}_\gamma(z) \T{e}^{ - \T{j} k_{\gamma x} x} \T{e}^{ - \T{j} k_{\gamma y} y}
\end{equation}
with
\begin{equation}
  k_{\gamma x} = k_{\T{i}x} + \kappa_{\gamma x},\quad k_{\gamma y} = k_{\T{i}y} + \kappa_{\gamma y}.
  \label{eq:kxky}
\end{equation}

Above ($+$) and below ($-$) the planes delimiting the extent of the periodic scatterer, the scattered field can be written in terms of a superposition of plane waves corresponding to Floquet harmonics
\begin{equation}
  \V{E}_\pm ^\T{sc} \left( x,y,z \right) = \sum\limits_\gamma \tilde{\V{e}}_{\{\pm,\gamma\}}^\T{sc} (z) \T{e}^{ - \T{j} k_{\gamma x} x} \T{e}^{ - \T{j} k_{\gamma y} y}.
  \label{eq:esc-exp}
\end{equation}
The function $\tilde{\V{e}}_{\{\pm,\gamma\}}^\T{sc} \left( z \right)$ associated with each scattered plane wave can be decomposed into two orthogonal polarizations (TE / TM or parallel / perpendicular) which are here denoted by unit vectors $\{\u{e}_p\}$. Performing the same form of factorization as in \eqref{eq:jper} to the scattered field~\eqref{eq:esc-exp} and separating polarizations~$p$, leads to a further expansion of the periodic scattered field
\begin{equation}
\label{eq:EscTot}
  \V{e}_\pm ^\T{sc} \left( \V{r} \right) = k \sqrt{\eta} \sum\limits_{\gamma, p} b_{\{\pm,\gamma, p\}} \V{u}_{\{\pm,\gamma, p\}} \left( \V{r} \right),
\end{equation}
where
\begin{equation}
  \V{u}_{\{\pm,\gamma, p\}} \left( \V{r} \right) = \u{e}_p \sqrt{\dfrac{1}{ k T_x T_y k_{\gamma z}}} \T{e}^{ - \T{j} \left( \kappa_{\gamma x} x + \kappa_{\gamma y} y \pm k_{\gamma z}z \right) }.
  \label{eq:bloch-1}
\end{equation}
are Bloch functions and 
\begin{equation}
  b_{\{\pm,\gamma, p\}} = \sqrt{\frac{T_x T_y k_{\gamma z}}{k\eta}}\u{e}_p\cdot\tilde{\V{e}}^\T{sc}_{\{\pm,\gamma\}} (z) \T{e}^{\pm\T{j}k_{\gamma z}z}.
  \label{eq:bloch-2}
\end{equation}
The particular normalization terms are selected such that the Bloch functions $\V{u}_{\{\pm,\gamma, p\}}$ are unitless and the coefficients $b_{\{\pm,\gamma, p\}}$ are related to the cycle mean power radiated per unit cell in the $\pm z$ direction via
\begin{equation}
\label{eq:Psc}
  P^\T{sc}_{\pm} = \dfrac{1}{2} \sum\limits_{\gamma ,p} \left| b_{\{ \pm ,\gamma ,p \}} \right|^2.
\end{equation}
In the language of microwave circuit theory~\cite{pozar2011microwave}, this choice makes the coefficients~$b_{\{ \pm ,\gamma ,p \}}$ represent outgoing power waves.

To connect the coefficients $b_{\{\pm,\gamma, p\}}$ to the periodic current~$\V{j}$, we write the scattered field $\tilde{\V{e}}_{\{\pm,\gamma\}}^\T{sc}$ in terms of the corresponding Fourier component of the current density and a dyadic Green's function \cite{kristensson2016scattering}
\begin{equation}
  \tilde{\V{e}}_{\{\pm,\gamma\}}^\T{sc} (z) = -\T{j}k\eta \int \V{g}_{\{\pm,\gamma\}}(z,z')\cdot\tilde{\V{\jmath}}_\gamma(z')\,\T{d}z'.
  \label{eq:esc-1}
\end{equation}
Above and below the scatterer, the dyadic Green's function can be factored as (\cf \cite[Eq. (10.49)]{kristensson2016scattering})
\begin{equation}
  \V{g}_{\{\pm,\gamma\}}(z,z') = \tilde{\V{g}}_{\{\pm,\gamma\}}\T{e}^{\mp\T{j}k_{\gamma z}z}\T{e}^{\pm\T{j}k_{\gamma z}z'}
\end{equation}
which reduces \eqref{eq:esc-1} to
\begin{equation}
\label{eq:etilde1}
  \tilde{\V{e}}_{\{\pm,\gamma\}}^\T{sc} (z) = -\T{j}k\eta \T{e}^{\mp\T{j}k_{\gamma z}z} \int \tilde{\V{g}}_{\{\pm,\gamma\}}\cdot\tilde{\V{\jmath}}_\gamma(z')\T{e}^{\pm\T{j}k_{\gamma z}z'}\T{d}z',
\end{equation}
where the longitudinal wavenumber is given by
\begin{equation}
  k_{\gamma z} = \sqrt{k^2 - k_{\gamma x}^2 - k_{\gamma y}^2}.
\end{equation} 
Expanding the periodic current density in a set of basis functions $\{\V{\psi}_\alpha\}$
\begin{equation}
  \V{j}(\V{r}) = \sum_{\alpha}I_\alpha \V{\psi}_\alpha(\V{r})
  \label{eq:j-exp}
\end{equation}
and combining \eqref{eq:jtilde}, \eqref{eq:etilde1}, \eqref{eq:bloch-1}, \eqref{eq:bloch-2} leads to the desired expression of $b_{\{ \pm ,\gamma ,p \}}$ in terms of the current expansion coefficients $\M{I}$,
\begin{equation}
\label{eq:BfromI}
  b_{\{\pm,\gamma, p\}} = - \sum_{\alpha}I_\alpha U^\alpha_{\{\pm,\gamma, p\}} = - \M{U}_{\{\pm,\gamma, p\}} \M{I}.
\end{equation}
The matrix~$\M{U}_{\{ \pm ,\gamma ,p \}}$ collects the projections of Bloch functions~\eqref{eq:bloch-1} onto basis functions~$\V{\psi}$, \ie, 
\begin{equation}
\label{eq:Umat}
  U^\alpha_{\{\pm,\gamma, p\}} = \dfrac{k \sqrt{\eta}}{2 } \int \V{u}_{\{\pm,\gamma, p\}}^* \left( \V{r} \right) \cdot\V{\psi}_\alpha(\V{r}) \,\T{d}\V{r}.
\end{equation}
This last relation assumes that $k_{\gamma z}$ is real valued (propagating Bloch modes) and utilizes the fact that the vectors $\{\u{e}_p\}$ have the eigenvalue property 
\begin{equation}
  \u{e}_{p}\cdot\tilde{\V{g}}_{\{\pm,\gamma\}} = - \dfrac{\T{j}}{2 k_{\gamma z}} \u{e}_{p}.
\end{equation}
For simplified notation in the remainder of the paper, the multi-index $m = \{\pm,\gamma,p\}$ is adopted to completely define the propagation direction, Floquet harmonic, and polarization of a given Bloch wave, \eg, $\M{U}_{\{ \pm ,\gamma ,p \}} = \M{U}_m$. 


To show that operator $\M{U}_m$ can be used to represent incident plane wave excitation, consider the electric field integral equation (EFIE) relating the periodic current density to an incident field distribution via an appropriate surface or volumetric impedance operator. Utilizing the expansion in \eqref{eq:j-exp} and applying Galerkin testing leads to the method of moments (MoM) representation of the EFIE
\begin{equation}
  \M{V} = \M{Z}\M{I}
  \label{eq:vzi}
\end{equation}
where $\M{Z}$ is the impedance matrix
and $\M{V}$ contains the projection of the periodic incident field onto the selected basis
\begin{equation}
V^\alpha = \int \V{e}^\T{i} \left( \V{r} \right) \cdot\V{\psi}_\alpha \left( \V{r} \right) \T{d}\V{r}.
\label{eq:valpha}
\end{equation}
Let the periodic part of incident field take the form similar to~\eqref{eq:EscTot}, \ie,
\begin{equation}
\label{eq:EincTot}
  \V{e}_\pm ^\T{i} \left( \V{r} \right) = k \sqrt{\eta} \sum\limits_m a_m \V{u}_m \left( \V{r} \right).
\end{equation}
Substitution into~\eqref{eq:valpha} and comparing with \eqref{eq:Umat} leads to 
\begin{equation}
\label{eq:VfromA}
\M{V} = 2 \sum\limits_m a_m \M{U}^\T{H}_m.
\end{equation}
Similarly to scattered power~\eqref{eq:Psc}, the cycle mean incident power passing through a rectangular patch of size~$T_x \times T_y$ which is normal to $z$ can be written as
\begin{equation}
\label{eq:Pinc}
  P^\T{i} = \dfrac{1}{2} \sum\limits_m \left| a_m \right|^2.
\end{equation}
Using \eqref{eq:vzi}, \eqref{eq:VfromA} and \eqref{eq:BfromI} yields a relation between the impedance matrix $\M{Z}$ and the elements of the matrix $\tilde{\M{S}}$
\begin{equation}
  \tilde{S}_{mm'} = \dfrac{b_m}{a_{m'}} = - 2 \M{U}_m \M{Z}^{-1}  \M{U}^\T{H}_{m'}.
\label{eq:s-from-z}
\end{equation}

The cycle mean scattered power per unit cell can be also written in terms of the radiation part of the impedance matrix or as the sum over all scattered powers, \ie,
\begin{equation}
  P_\T{sc} = \frac{1}{2}\M{I}^\T{H}\M{R}\M{I} = \frac{1}{2}\M{I}^\T{H}\left(\sum_m\M{U}_m^\T{H}\M{U}_m\right)\M{I}.
  \label{eq:psc-r}
\end{equation}
Because the above expression holds for all currents $\M{I}$, it holds that the two matrices within the quadratic forms are equal. The impedance-based characteristic mode eigenvalue problem in \eqref{eq:gep-z} can be rearranged to
\begin{equation}
  \M{Z}^{-1}\M{R}\M{I}_n = \frac{1}{1+\T{j}\lambda_n}\M{I}_n
\end{equation}
and substituting the alternate representation of the matrix $\M{R}$ from \eqref{eq:psc-r} into the above expression gives
\begin{equation}
\M{Z}^{-1}\left(\sum_{m}\M{U}_{m}^\T{H}\M{U}_{m}\right)\M{I}_n = \frac{1}{1+\T{j}\lambda_n}\M{I}_n.
\end{equation}
Rearranging the above expression and left multiplying with $\M{U}_m$ leads to
\begin{equation}
\sum_{m'}\M{U}_{m}\M{Z}^{-1}\M{U}_{m'}^\T{H} \M{U}_{m'}\M{I}_n \\= \frac{1}{1+\T{j}\lambda_n}\M{U}_{m}\M{I}_n
\end{equation}
which, by \eqref{eq:s-from-z} and \eqref{eq:BfromI} reduces to
\begin{equation}
  \sum_{m'}\tilde{S}_{mm'}b_{n,m'} = -\frac{2}{1+\T{j}\lambda_n}b_{n,m}.
\end{equation}
Collecting this form of equation for all indices $m$ yields the eigenvalue problem in \eqref{eq:new-ep-st} and the relation between impedance- and scattering-based eigenvalues in \eqref{eq:lambda-t}.

\bibliographystyle{ieeetr}
\bibliography{refs.bib}


\begin{IEEEbiography}[{\includegraphics[width=1in,height=1.25in,clip,keepaspectratio]{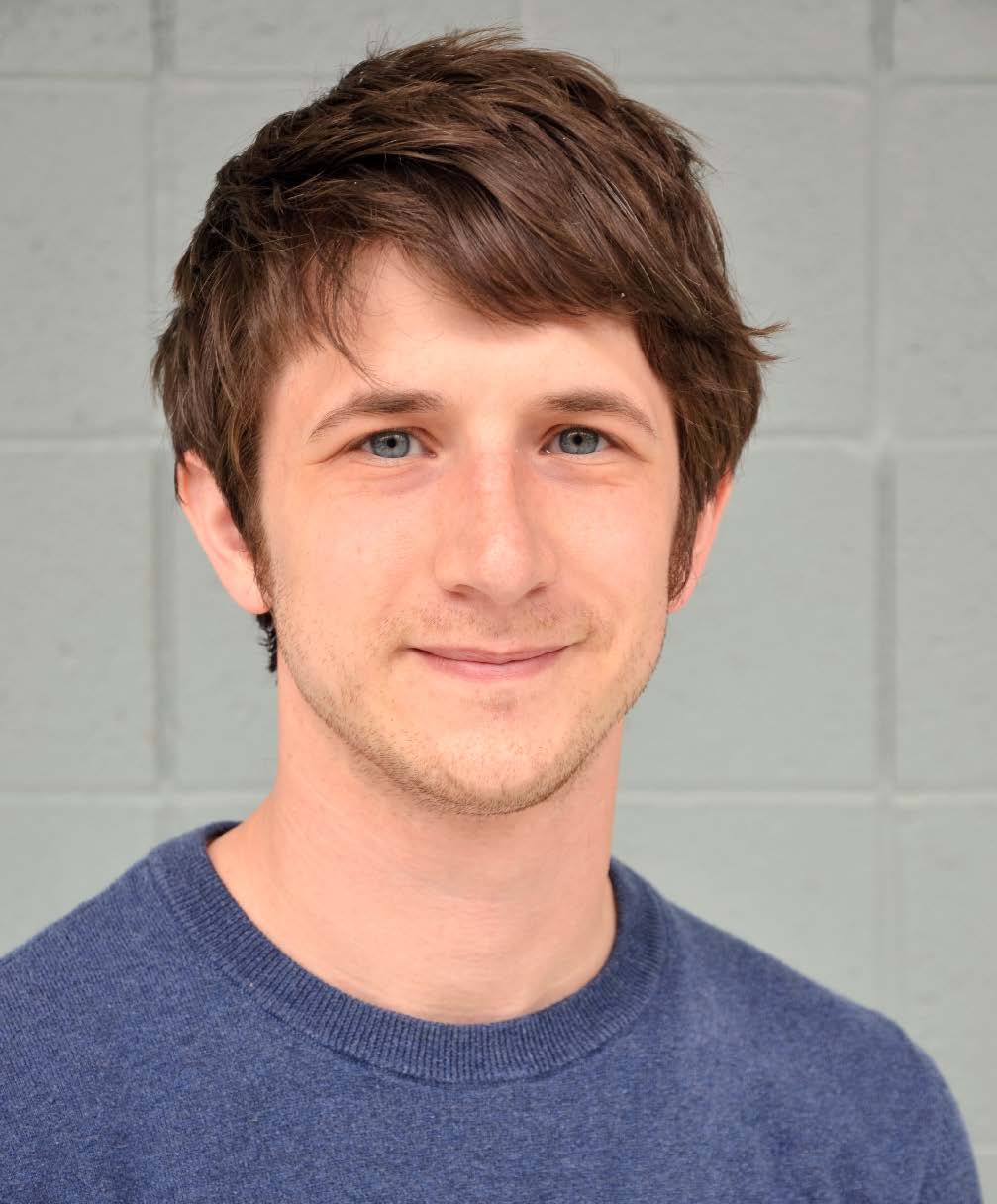}}]{Kurt Schab}
	(S'09, M'16) is an Assistant Professor of Electrical Engineering at Santa Clara University, Santa Clara, CA USA. He received the B.S. degree in electrical engineering and physics from Portland State University in 2011 and the M.S. and Ph.D. degrees in electrical engineering from the University of Illinois at Urbana-Champaign in 2013 and 2016, respectively.  From 2016 to 2018 he was a Postdoctoral Research Scholar at North Carolina State University in Raleigh, North Carolina.  His research focuses on the intersection of numerical methods, electromagnetic theory, and antenna design.  
\end{IEEEbiography}

\begin{IEEEbiography}[{\includegraphics[width=1in,height=1.25in,clip,keepaspectratio]{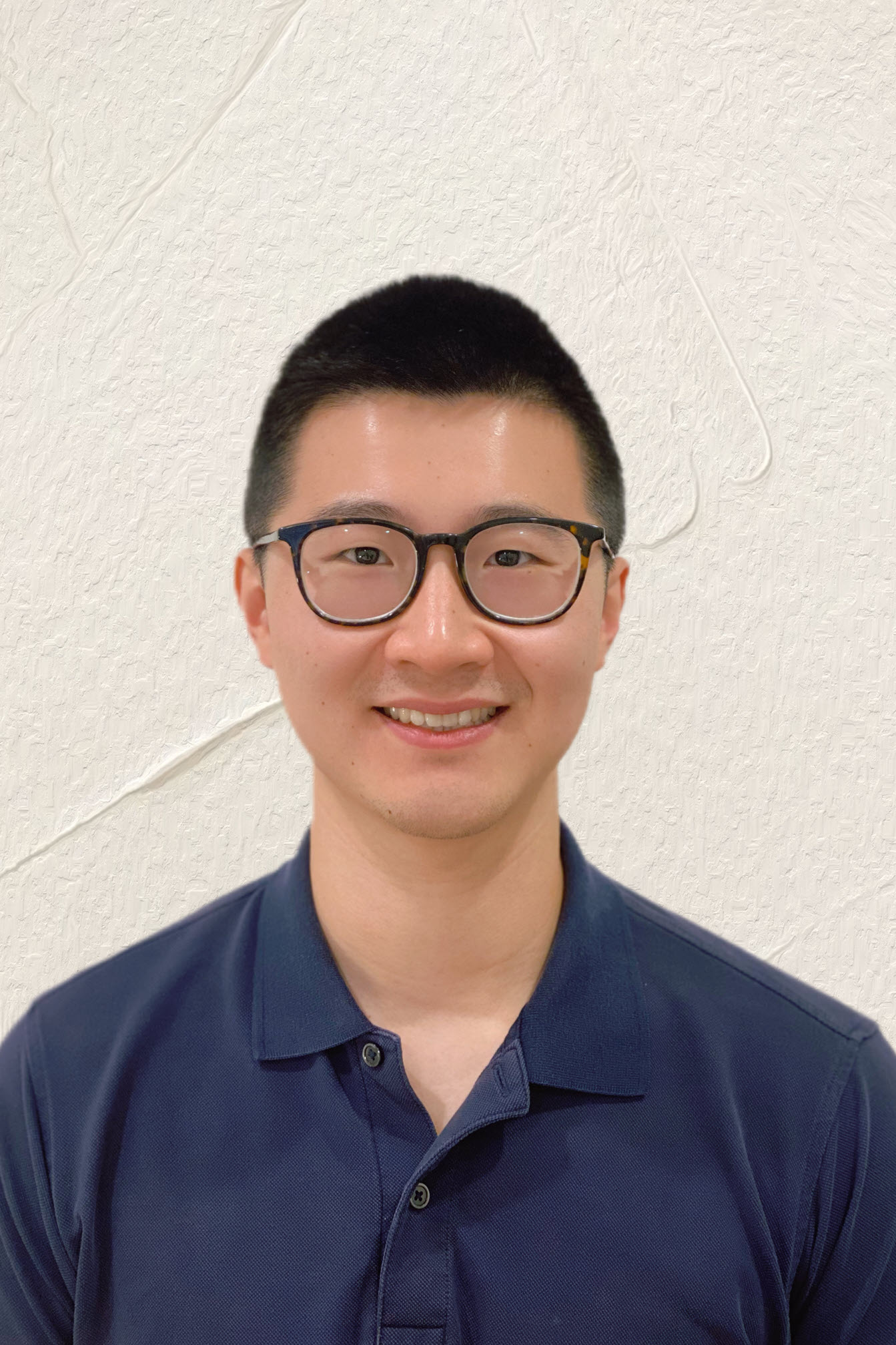}}]{Frederick Chen}
	(S'13) received the B.S. degree in Electrical Engineering and the M.S. degree in Electrical and Computer Engineering from the Georgia Institute of Technology, Atlanta, Georgia, USA, in 2013 and 2017, respectively.
	
	He is currently a Doctoral Student in the Electrical Engineering department of Santa Clara University, Santa Clara, California, USA. Before joining Santa Clara University, he spent several years in the industry and the US military. His research interests are in computational electromagnetics, metamaterials, and electromagnetic applications.
\end{IEEEbiography}

\begin{IEEEbiography}[{\includegraphics[width=1in,height=1.25in,clip,keepaspectratio]{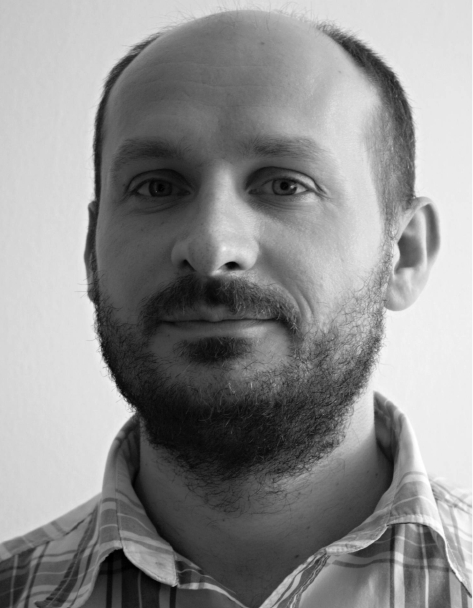}}]{Lukas Jelinek}
	received his Ph.D. degree from the Czech Technical University in Prague, Czech Republic, in 2006. In 2015 he was appointed Associate Professor at the Department of Electromagnetic Field at the same university.
	
	His research interests include wave propagation in complex media, electromagnetic field theory, metamaterials, numerical techniques, and optimization.
\end{IEEEbiography}

\begin{IEEEbiography}[{\includegraphics[width=1in,height=1.25in,clip,keepaspectratio]{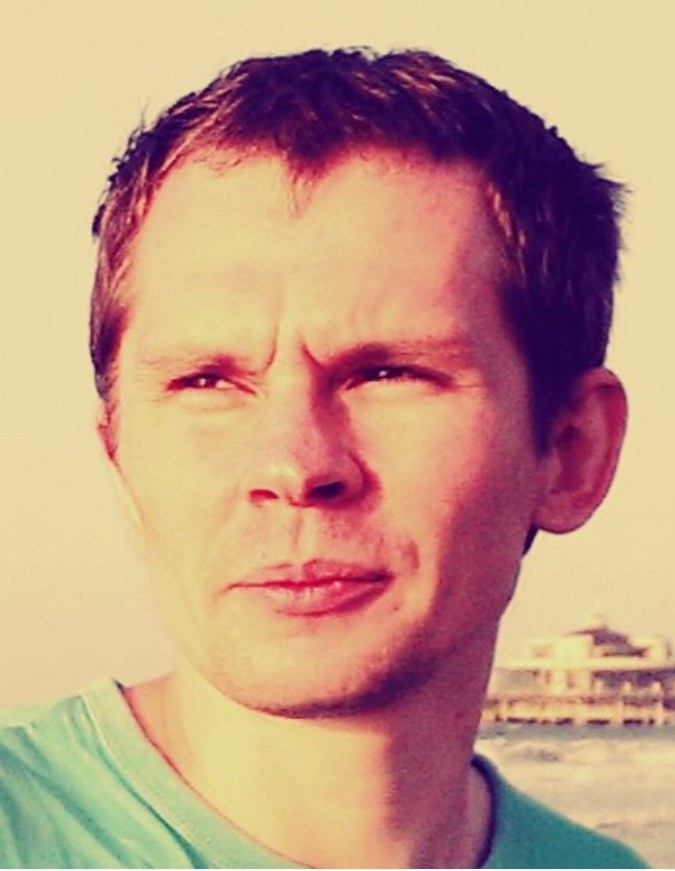}}]{Miloslav Capek}
	(M'14, SM'17) received the M.Sc. degree in Electrical Engineering 2009, the Ph.D. degree in 2014, and was appointed Associate Professor in 2017, all from the Czech Technical University in Prague, Czech Republic.
	
	He leads the development of the AToM (Antenna Toolbox for Matlab) package. His research interests are in the area of electromagnetic theory, electrically small antennas, antenna design, numerical techniques, and optimization. He authored or co-authored over 100~journal and conference papers.
	
\end{IEEEbiography}

\begin{IEEEbiography}[{\includegraphics[width=1in,height=1.25in,clip,keepaspectratio]{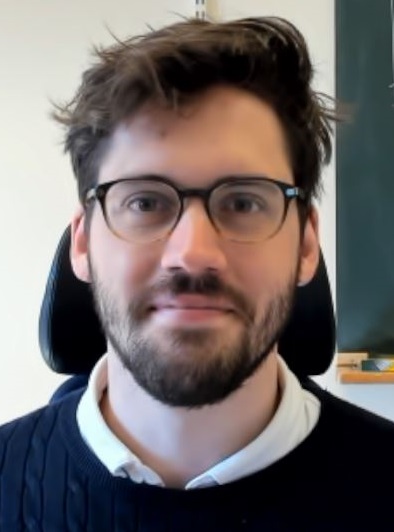}}]{Johan Lundgren}
	(M'22) is a postdoctoral researcher at Lund University. he received his M.Sc degree in engineering physics 2016 and Ph.D. degree in Electromagnetic Theory in 2021 all from Lund University, Sweden. 
	
	His research interests are in electromagnetic scattering, wave propagation, computational electromagnetics, functional structures, meta-materials, inverse scattering problems, imaging, and measurement techniques.
	
\end{IEEEbiography}

\begin{IEEEbiography}[{\includegraphics[width=1in,height=1.25in,clip,keepaspectratio]{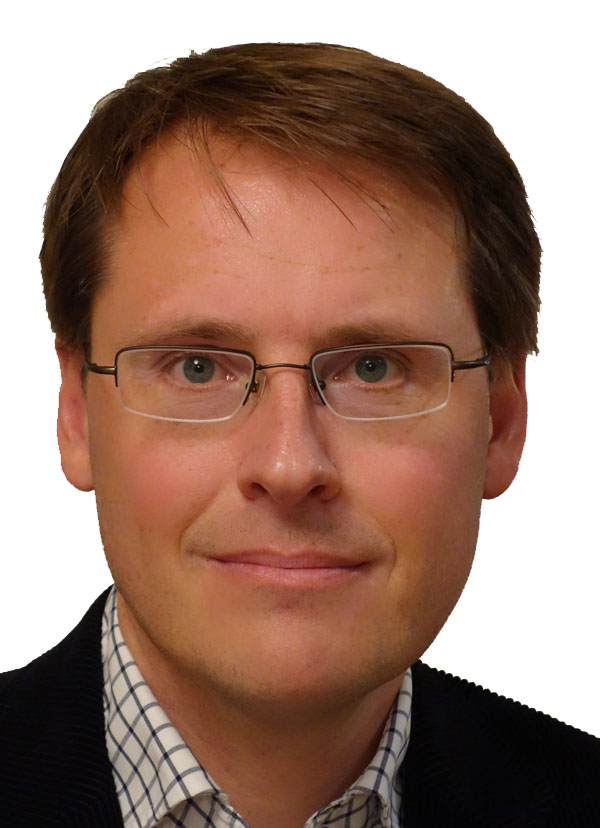}}]{Mats Gustafsson}
	received the M.Sc. degree in Engineering Physics 1994, the Ph.D. degree in Electromagnetic Theory 2000, was appointed Docent 2005, and Professor of Electromagnetic Theory 2011, all from Lund University, Sweden.
	
	He co-founded the company Phase holographic imaging AB in 2004. His research interests are in scattering and antenna theory and inverse scattering and imaging. He has written over 100 peer reviewed journal papers and over 100 conference papers. Prof. Gustafsson received the IEEE Schelkunoff Transactions Prize Paper Award 2010, the IEEE Uslenghi Letters Prize Paper Award 2019, and best paper awards at EuCAP 2007 and 2013. He served as an IEEE AP-S Distinguished Lecturer for 2013-15.
\end{IEEEbiography}

\end{document}